\documentclass[prl,twocolumn,showpacs]{revtex4-1}
\usepackage{graphicx}
\usepackage{dcolumn}
\usepackage{bm}
\usepackage{color}

\begin{document}

\title{Oblique Half-Solitons and their Generation in Exciton-Polariton Condensates}
\author{H. Flayac, D.D. Solnyshkov, and G. Malpuech}
\affiliation{Clermont Universit\'{e}, Universit\'{e} Blaise Pascal, LASMEA, BP10448, 63000 Clermont-Ferrand, France}
\affiliation{CNRS, UMR6602, LASMEA, 63177 Aubi\`{e}re, France}

\begin{abstract}
We describe oblique half-solitons, a new type of topological defects in a two dimensional spinor Bose Einstein condensate. A realistic protocol based on the optical spin Hall effect is proposed toward their generation within an exciton-polariton system.
\end{abstract}

\maketitle

\emph{Introduction.} Exciton-polaritons (polaritons) are hybrid
quasi-particles that share the properties of both light and matter. On one hand,
 their excitonic part allows them to interact both with
themselves and with the surrounding phonon bath. In microcavities, this peculiarity leads to
efficient relaxation processes towards the ground state and to the formation
of a Bose-Einstein condensate (BEC)\cite{BECPolaritons} with extended
spatial coherence\cite{Wertz}. On the other hand, the photonic part of polaritons
attributes them a very small effective mass which allows condensation even
at room temperature\cite{RoomTemp}.

While the polariton BEC has indeed been obtained experimentally, the superfluidity,
namely the appearance of a dissipationless flow, was still an issue. Strong progress has been made in 2009 when the behavior
of a flow of polaritons past an obstacle was analyzed\cite{AmoNat,AmoNatPhys}. One of the fascinating
properties of a BEC is the possible creation of topological defects by
a perturbation. In spinor polariton condensates, elementary excitations are peculiar: indeed, two exciton spin projections ($\sigma^\pm=\pm1$) can
efficiently couple with light. Thus the wave function (order parameter) of a polariton BEC is
no more scalar but vectorial. In such a
framework elementary topological defects carry half-integer angular momentum and are called
half-vortices (HV). These objects were predicted theoretically\cite{RuboHV}
in a thermodynamic limit, analyzed in presence of various (effective)
magnetic fields\cite{RuboPS, FlayacTETM, RuboMF} and observed
experimentally\cite{LagoudakisHV} as a result disorder in
microcavities.

A soliton is another example of topological defect that can
basically be seen as a 1D analogue of a vortex. It is characterized
by a maximum phase shift of $\pi$ (dark soliton), on the scale of the
healing length $\xi$ of the condensate, accompanied by a ditch in the density profile.
In 2D, a new type of extended defect has been recently
described: oblique dark solitons (ODS)\cite{ObliqueDarkSolitons}, which were
shown to be persistent or rather only convectively unstable\cite
{PitaevskiiS}. They can be generated if a supersonic condensate flow hits a
defect larger than $\xi$\cite{ObliqueDarkSolitons, CiutiVortex}. These
predictions have been recently verified in a polaritonic system\cite
{AmoSoliton}. Polaritons can also exhibit bright solitons\cite{Skryabin}, but these are not in the focus of the present Letter.
On the other hand, solitons in spinor 1D
 condensates\cite{DarkSolitonSpinor,FamilyMatterWaves} and oblique solitons in
spinor 2D systems have also been theoretically considered\cite
{SpinorDarkSoliton2D}. In one dimension, many
possible configurations were described, depending on the strength
and type of the particle interactions (repulsive or attractive).
In particular, a solution where the kink
lies in only one component was reported: the dark-antidark soliton.
This kind of defect reminds the HV configuration and we will see why it
deserves to be called a "half-soliton"\cite{VolovikHS}.

In the present work we discuss the generation of oblique dark
half-solitons (ODHS), in a spinor BEC. The paper is
organized as follows. First, we analyze the 1D half-soliton by
analogy with the half-vortex. Second, we introduce the ODHS.
Third, we propose two experimental configurations leading to their formation,
capitalizing on the specificities of polariton condensates. The first
protocol is using a defect acting only on one circular component. The second
one is using the TE-TM splitting present in planar microcavities.

\emph{The 1D half-soliton.} A two components polariton condensate at $0$K can be described by a spinor
Gross-Pitaevskii equation\cite{Shelykh2006}. Assuming first a
parabolic dispersion with an effective mass $m^{\ast }$ and an
infinite lifetime of the particles, one has
\begin{eqnarray}
i\hbar \frac{{\partial {\Psi _\sigma }}}{{\partial t}} =  - \frac{{{\hbar ^2}}}{{2{m^ * }}}\Delta {\Psi _\sigma } + \left( {{\alpha _1}{{\left| {{\Psi _\sigma }} \right|}^2} + {\alpha _2}{{\left| {{\Psi _{ - \sigma }}} \right|}^2}} \right){\Psi _\sigma }
\end{eqnarray}
Time-independent solutions are found upon expressing the BEC wave functions as ${%
\Psi _{\pm }}\left( {\mathbf{r},t}\right) ={\psi _{\pm }}\left( \mathbf{r}%
\right) \exp \left( {-i\mu t}\right) $, where $\mu =(\alpha _{1}+\alpha
_{2})n_{0}$ is the chemical potential related to the density of the
homogeneous condensate $n_{0}=n_{0+}/2=n_{0-}/2$. Indeed, for a realistic
polariton BEC, the interactions between the particles of opposite spin are weak and attractive: $\alpha
_{2}\simeq -0.2\alpha _{1}$ which leads to a linearly polarized ground
state. When looking for a single-kink solution such as a vortex or a
soliton, it is convenient to discuss the asymptotic behavior. We assume that
far away from the defect the condensate density is unperturbed.

\begin{figure}[tbp]
\includegraphics[width=0.40\textwidth,clip]{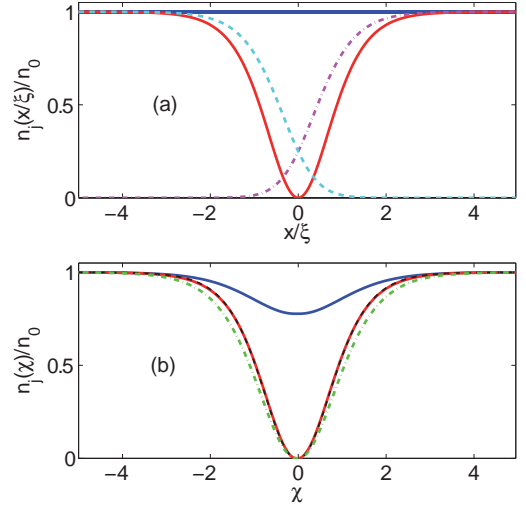}
\caption{(Color online) (a) HS density profiles scaled to $n_{0}$ and $%
\protect\xi $. The solid blue and red curves represent the $n_{+}$ and $%
n_{-}$ density profiles while the dashed-dotted purple and dashed cyan curves show $n_{x}$
and $n_{y}$ respectively. (b) ODHS density profiles normal to its axis. The solid
blue and red curves show numerical profiles, the dashed black curve is the
perturbative solution described in the text and the dashed dotted green curve
show the scalar soliton solution. The parameters are $a=10$, $U=3.5$, $\Lambda _{1}=1$ and $%
\Lambda _{2}=-0.2$.
}
\label{Fig1}
\end{figure}

The order parameter can be written in two
distinct representations\cite{RuboHV,FlayacTETM}. On the basis of circular
polarizations discussed previously, each component possesses its own phase $%
\theta _{\pm }(\mathbf{r})$ and $\left( {{\psi _{+}},{\psi _{-}}}\right) =%
\sqrt{{n_{0}}}/2\left( {{{\mathop{\rm e}\nolimits}^{i{\theta _{+}}}},{{%
\mathop{\rm e}\nolimits}^{i{\theta _{-}}}}}\right) $. In the linear
polarization basis, the polarization angle $\eta (\mathbf{r})$ and the
global phase of the condensate $\varphi (\mathbf{r})$ can be separated: $%
\left( {{\psi _{x}},{\psi _{y}}}\right) =\sqrt{{n_{0}}}\left( {{{\mathop{\rm
e}\nolimits}^{i\varphi }}\cos \left( \eta \right) ,{{\mathop{\rm e}\nolimits}%
^{i\varphi }}\sin \left( \eta \right) }\right) $. The transformation from
one to another is obtained via ${\psi _{\pm }}=\left( {{\psi _{x}}\mp i{\psi
_{x}}}\right) /\sqrt{2}$ and ${\theta _{\pm }}=\varphi \mp \eta $. In the
latter form, half-vortices are characterized by two half-integer winding
numbers for both the phase and the polarization which lead to integer
shifts of $\pi $ around their core. Similar considerations can be applied to
solitons. Indeed, in one dimension the scalar dark soliton solution\cite
{PitaevskiiBook} is simply given by ${\psi _{S}\left( x\right) }=\sqrt{{n_{0}%
}}\tanh \left( x\right) $ and its phase is a Heaviside function of amplitude
$\pi $. In the circular polarization basis and, in the simplest case where $%
\alpha _{2}=0$, a half-soliton (HS) is nothing but a usual soliton appearing
in one component (let's say $\psi _{-}$) while the other remains
homogeneous. Thus, the associated order parameter reads $\left( {\psi
_{+}^{HS},\psi _{-}^{HS}}\right) =\sqrt{{n_{0}}/2}\left( {1,\tanh \left(
x\right) }\right) $. Rewriting the latter on the linear polarization basis
leads to $\left( {\psi _{x}^{HS},\psi _{y}^{HS}}\right) =\sqrt{{n_{0}}}%
/2\left( {1+\tanh \left( x\right) ,i-i\tanh \left( x\right) }\right) $.
Looking at asymptotic forms, one easily obtain
\begin{eqnarray}
\psi _{x}^{HS}\left( {+\infty }\right)  &=&\sqrt{{n_{0}}}{e^{2ih\pi }}\cos
\left( {2s\pi }\right)   \nonumber \\
\psi _{y}^{HS}\left( {+\infty }\right)  &=&\sqrt{{n_{0}}}{e^{2ih\pi }}\sin
\left( {2s\pi }\right)  \\
\psi _{x}^{HS}\left( {-\infty }\right)  &=&\sqrt{{n_{0}}}{e^{ih\pi }}\cos
\left( {s\pi }\right)   \nonumber \\
\psi _{y}^{HS}\left( {-\infty }\right)  &=&\sqrt{{n_{0}}}{e^{ih\pi }}\sin
\left( {s\pi }\right)   \nonumber
\end{eqnarray}
with $h$ and $s$ half-integer numbers which can be seen as topological
charges. Basic HS appear for $(h,s)=\{(\pm 1/2,\pm 1/2),(\pm 1/2,\mp 1/2)\}$
and their phase and polarization angle are ranging from 0 to $\pi /2$ with a
fully circularly polarized center just like the half-vortex core. This
topological defect can also be seen as a domain wall with respect to x- and
y-polarized particles. A plot of the HS density profiles ($n_{j}={\left| {%
\psi _{j}^{HS}}\right| ^{2}}$, $j=\pm,x,y$) is proposed in Fig.\ref{Fig1}(a).

\emph{The oblique half-soliton.} For a 2D condensate it was shown\cite
{ObliqueDarkSolitons,PitaevskiiS,SpinorDarkSoliton2D} that stationary
extended oblique dark solitons can occur both in the scalar and spinor systems
despite their instability, because the latter is only convective.
They appear in a perturbed fluid at supersonic velocities as merging
vortices arranged in "trains" ("streets").

Let us now focus on the possibility
of creating ODHS. First of all, it is clear that
in the case where the two components of the spinor BEC do not interact ($%
\alpha _{2}=0$), if they are initially equally populated, a
perturbation in only one of the components will lead to the formation of half-integer
excitations. Next, what happens if the interaction is no longer negligible?
To answer this question we turn back to Eqs.(1) and (2), rescale them, look
for stationary solutions and switch to the hydrodynamic picture of Ref.%
\onlinecite{SpinorDarkSoliton2D} where the phase of each components is
expressed by means of their stationary and irrotational velocity fields via $%
\mathbf{{v_{\pm }}}\left( {\mathbf{r}}\right) =\hbar /m^{\ast }%
\boldsymbol{\nabla} {\theta _ \pm }\left( {\mathbf{r}}\right) $, with $%
\mathbf{r}=\left( {x,y}\right) $. We look for oblique solutions that depend
only on the tilted coordinate $\chi =(x-ay)/\sqrt{1+{a^{2}}}$ which leads to
the following set of equations
\begin{eqnarray}
\left( {n{{_{\sigma}^{\prime }}^{2}}/4-{n_{\sigma}}n_{\sigma}^{\prime \prime }/2}\right)
&+&2n_{\sigma}^{2}\left( {{\Lambda _{1}}{n_{\sigma}}+2{\Lambda _{2}}{n_{-\sigma}}}\right)
\nonumber \\
&=&\left( {q+2\mu }\right) n_{\sigma}^{2}-qn_{0}^{2}
\end{eqnarray}
where $\Lambda _{1,2}=\alpha _{1,2}/(\alpha _{1}+\alpha _{2})$ and $%
q=U^{2}/(1+a^{2})$ ($U$ the velocity of the flow). This system has to be
solved numerically, but let us start with some simple arguments. The density profile of an integer
oblique dark soliton\cite{ObliqueDarkSolitons} is given by ${n _{ODS}}%
=1-\left( {1-q/\mu }\right) {\mathop{\rm sech}\nolimits}{\left[ {\chi \sqrt{%
\mu -q}}\right] ^{2}}$ with $\mu =\Lambda _{1}n_{0}$. Now, for the case of
the ODHS, the density notch in the $\sigma ^{-}$ component, that contains the
defect, is seen as an external potential by the initially unperturbed $\sigma ^{+}$
component, because of the interactions between the particles of different spins. We suppose that the $\sigma ^{+}$ component fits the
shape of this potential which is nothing but $\Lambda _{2}n_{-}$. Then, this
perturbation creates in turn a potential for the $\sigma ^{-}$ component given
by $-\Lambda _{2}n_{+}=-\Lambda _{2}^{2}n_{-}$. Therefore, the density profile is
modified as ${{\widetilde{n}_{-}}}\leftarrow \left( {{\Lambda _{1}}{n_{-}}%
-\Lambda _{2}^{2}{n_{-}}}\right) /{\Lambda _{1}}$.
Iterating this procedure leads to a geometric series and to a renormalization of the interaction constant $\widetilde{\Lambda }_{1} \leftarrow {\Lambda _1} - \Lambda _2^2/\left( {{\Lambda _1} - {\Lambda _2}} \right)$. Consequently, the
ODHS solution is approximated by
\begin{equation}
{n _{ODHS}}=1-\left( {1-q/\widetilde{\mu }}\right) {\mathop{\rm sech}%
\nolimits}{\left[ {\chi \sqrt{\widetilde{\mu }-q}}\right] ^{2}}
\label{ODHSSol}
\end{equation}
with $\widetilde{\mu }=\widetilde{\Lambda }_{1}n_{0}$. In this description, the sound
velocity is changed like ${c_{s}}\rightarrow {\widetilde{c}_{s}}=\sqrt{%
\widetilde{\mu }/{m^{\ast }}}$ and the healing length like $\xi \rightarrow
\widetilde{\xi }=\hbar /\sqrt{2{m^{\ast }}\widetilde{\mu }}$. In the case
where $\Lambda _{2}<0(>0)$, which corresponds to an attractive (repulsive)
interaction, $c_{s}$ is slightly increased (decreased) and inversely for $%
\xi $. The component without a soliton obviously presents
a minimum (maximum). This argumentation is compared to direct numerical solutions of Eqs.
(4) and (5) in the Fig.\ref{Fig1}(b)  showing a remarkable accuracy for small
values of $\Lambda _{2}$.

\emph{Generation of ODHS in a polariton fluid.} Bringing an external perturbation in a quantum
fluid is an efficient way to nucleate topological defects in a controlled manner. A relevant experimental setup relies on the acceleration of a superfluid
above the sound velocity past an obstacle larger than $\xi$, where the
breakdown of superfluidity allows the elementary excitations to develop. The
defect is a potential barrier resulting from an off-resonance laser
beam. A planar microcavity is a system that offers a high degree of control
on the velocity and the density of 2D spinor BECs as discussed in
Refs.\onlinecite{Tejedor,AmoSoliton}.

Now, to describe more accurately the polariton condensate, we take into
account the real non parabolic dispersion of the particles via a set of four coupled spin-dependent equations for the photonic $\Psi _{\pm }^{ph}$ and
excitonic $\Psi _{\pm }^{ex}$ fields respectively. A resonant pumping
situation is considered with photon lifetime of 100 ps.
\begin{eqnarray}
i\hbar \frac{{\partial \Psi _{\pm }^{ph}}}{{\partial t}} &=&-\frac{{{\hbar
^{2}}}}{{2{m_{ph}^{\ast }}}}\Delta \Psi _{\pm }^{ph}+D_{\pm }{\Psi _{\pm
}^{ph}}+\frac{\Omega _{R}}{2}\Psi _{\pm }^{ex}  \nonumber \\
&-&\frac{{i\hbar }}{{2{\tau _{ph}}}}\Psi _{\pm }^{ph}+{P_{\pm }}+\beta {%
\left( {\frac{\partial }{{\partial x}}\mp i\frac{\partial }{{\partial y}}}%
\right) ^{2}}\Psi _{\mp }^{ph} \\
i\hbar \frac{{\partial \Psi _{\pm }^{ex}}}{{\partial t}} &=&-\frac{{{\hbar
^{2}}}}{{2{m_{ex}^{\ast }}}}\Delta \Psi _{\pm }^{ex}+\frac{\Omega _{R}}{2}%
\Psi _{\pm }^{ph}  \nonumber \\
&-&\frac{{i\hbar }}{{2{\tau _{ex}}}}{\Psi _{\pm }^{ex}}+\left( {{\alpha _{1}}%
{{\left| {\Psi _{\pm }^{ex}}\right| }^{2}}+{\alpha _{2}}{{\left| {\Psi _{\mp
}^{ex}}\right| }^{2}}}\right) \Psi _{\pm }^{ex}
\end{eqnarray}
The new quantities that appear are a quasi-resonant \emph{cw} pumping which is
taken to be linearly polarized $P_{+}=P_{-}$, the strong
coupling between exciton and photons with a Rabi splitting $\Omega _{R}=10$ meV  and the
photonic TE-TM splitting\cite{Review} of strength $\beta $ for which we
assume a quadratic dependence over the wave vector\cite{FlayacTETM}. We
assume that the potential barrier we consider $D_{\pm }$
can affect independently each circularly polarized  photonic component.

In the numerical experiments that follow we reproduce the protocol described
in Refs.\onlinecite{CiutiVortex,AmoSoliton} involving an immobile defect and
a moving fluid. The pump laser is tuned slightly above
the bare polariton branch ($+0.3$ meV) to remain resonant despite the interaction-induced blue shift. The spot (15 $\mu$m upstream from the defect) is bar shaped ($x \times y=(10\times 100)\mu m^2$). The fluid propagates from
the left to the right with a wave vector $k_{p,x}=1\mu$m$^{-1}$, low
enough to avoid parametric instabilities, and collides against a $6$ $\mu$m large
circular defect. Defining the sound velocity is far from trivial so far
as the local pumping and propagation effects induce density fluctuations and its general exponential decrease.
 The sound
velocity ${c_{s}}=\sqrt{\mu /{m^{\ast }}}$ of an unperturbed flow is expected to decrease like $\sqrt {\exp \left( { - {m^*}d/\hbar k{\tau _{ph}}} \right)}$ where $m^{\ast }\sim
2m_{ph}^{\ast }$ and $d$ is the distance from the pump. Strictly speaking, $c_s$ has to be defined locally and separately for each components. We choose a
pseudospin representation, which is relevant to provide information about
both components of the order parameter in a single picture. We recall that
the pseudospin $\mathbf{S}\left( {\mathbf{r},t}\right) $ is a 3D vector lying on a Poincar\'{e} sphere. Its components are
defined by
\begin{eqnarray}
{S_{x}} &=&\Re \left( {\psi _{+}^{ph}\psi _{-}^{ph\ast }}\right)   \nonumber
\\
{S_{y}} &=&\Im \left( {\psi _{+}^{ph\ast }\psi _{-}^{ph}}\right)  \\
{S_{z}} &=&{{\left( {n_{+}^{ph}-n_{-}^{ph}}\right) }\mathord{\left/
{\vphantom {{\left( {n_ + ^{ph} - n_ - ^{ph}} \right)} 2}} \right.
\kern-\nulldelimiterspace}2}  \nonumber
\end{eqnarray}
normalized to unity. $S_{x,y}$ define linearly polarized
states while $S_{z}$ is the degree of circular polarization of the
condensate\cite{Review}. Below, we propose two different experimental configurations
 for the generation of ODHS.

\begin{figure}[tbp]
\includegraphics[width=0.45\textwidth,clip]{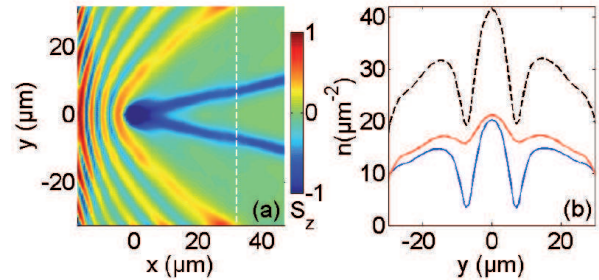}\newline
\caption{(a) Pseudospin representation of the order parameter
at $M\sim1.8$. The colormap shows the $S_{z}$ component (degree of
circular polarization). (b) Density slice displaying $n_{+}$ (blue curve) where the soliton appears, $n_{-}$
(red curve) with a minimum at the soliton position and the sum of the two (dashed black curve). The dashed white lines indicate the slice position.}
\label{Fig2}
\end{figure}

The first alternative is to find a way to perturb only one component. This can be done for polaritons where the defect potential could be optically created by a circularly polarized pulse\cite{Amo2010}. This scheme is however far from being ideal since it would bring of lot of unwanted perturbations to the system and it should rather be seen as a model experiment.  In this framework, we impose $D_{+}=20$ meV, $D_{-}=0$,  and $\alpha _{2}=-0.2\alpha _{1}$. The $S_z(x,y)$ component of the corresponding stationary flow regime, is presented in Fig.\ref{Fig2}(a) for a Mach number $M\sim1.8$. The generated pair of oblique solitons is left circularly ($\sigma_-$) polarized and with the help of a density slice normal to the flow Fig.\ref{Fig2}(b) we are able to undoubtedly identify an ODHS.

\begin{figure}[tbp]
\includegraphics[width=0.45\textwidth,clip]{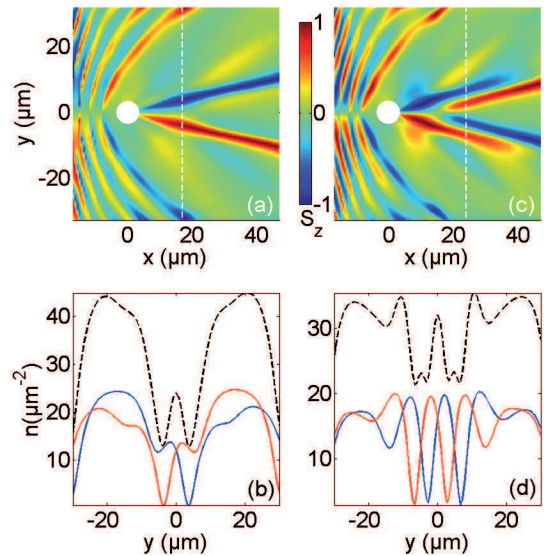}
\caption{Same representations as Fig.\ref{Fig2}. ODHS pair(s) in the presence of the TE-TM splitting. (a)-(b) $\beta k_{p,x}^2=0.02$ meV and (c)-(d) $\beta k_{p,x}^2=0.04$ meV.}
\label{Fig3}
\end{figure}

Our second proposal is, on the other hand, completely realistic. The impenetrable defect is restored in both components and we suggest to
benefit from the spatial separation brought by the $\mathbf{k}$-dependent
TE-TM splitting. The strength of the induced
effective magnetic field $\boldsymbol{\Omega}_{LT}$ is given by $\beta ={%
\hbar ^{2}}\left( {m_{l}^{-1}-m_{t}^{-1}}\right) /4$ where $%
m_{l,t}$ are the effective masses of TM and TE polarized particles
respectively. The energy splitting between the two latter eigen modes
is $\Delta _{LT}=\beta k_{p,x}^{2}=0.02$ meV for $m_{t}=5\times 10^{-5}m_{0}$
and $m_{l}=0.95m_{t}$ ($m_{0}$ is the free electron mass). Moreover, the
field makes a double angle with respect to the direction of propagation and
is thus symmetric with respect to the $x$-axis. The latter leads to the
so-called optical spin Hall effect: the precession of the
pseudospin around $\boldsymbol{\Omega}_{LT}$\cite{OSHEth,OSHEexp}. In our numerical setup, the linearly $x$-polarized
injected particles are initially moving along the $x$-direction so no
polarization separation is expected until the fluid reaches the obstacle.
Indeed, at this precise moment, the flow is split into two parts moving in oblique directions.
These two parts
start to precess about the effective field in opposite directions. Particles
going to the left gain a $\sigma^+$ component. Particles going to the right gain
a $\sigma^-$ component. This antisymmetry results in the formation of an oblique soliton in
one circular component, and not in the other, which is an ODHS. The figure \ref{Fig3}(a)
shows the resulting polarization pattern achieved in the steady
state: A pair of oblique dark solitons of
opposite circularity is formed behind the obstacle.
It should also be noted that the \~{C}erenkov radiations exhibit a remarkable polarization separation upstream from the defect. The figure \ref{Fig3}(b) shows a
density slice of each circular component that reveals the
ODHS pair. The profiles are qualitatively similar to the one obtained for the ideal
case presented on the figure \ref{Fig1}(b). The \ref{Fig3}(c) shows the results of a
similar calculation, but for a splitting twice larger. One can
clearly see the formation and disappearance of a pair of ODHS close to the
defect, and, further away, the formation of a second one of opposite
circularity. The figure \ref{Fig3}(d) shows the density when the four ODHS
are simultaneously present. This behavior is due to the TE-TM splitting
which couples the circular components and therefore couples ODHS of opposite
circularity.
To finish, we make the following remark:
In the transient regime, before the steady state is established or at lower fluid velocities,
we have clearly observed the generation of HVs, and it will be the topic of a
future work.

\emph{Conclusions.} We have reported a new type of half-integer topological
defect that occurs in two-components 2D BEC: the oblique dark
half-soliton. We have shown that an exciton-polariton fluid is well-suited for the study of such defects.
We have made a realistic proposal, based on the
optical spin Hall effect, which should lead to their experimental observation.
Besides, the setup we suggest should also allow a controlled nucleation of
half-vortices, which was never reported so far.


\end{document}